\documentclass[a4paper,aps,pra,reprint,superscriptaddress, floatfix,twocolumn]{revtex4-1}
\usepackage{amsmath}
\usepackage{verbatim}
\usepackage{graphicx}
\usepackage{physics}
\usepackage{color}
\usepackage[normalem]{ulem}
\usepackage{amsmath,amsfonts,amssymb,graphicx,color,times,bbm,psfrag,dsfont,slashed,bm}
\usepackage[colorlinks=true,citecolor=blue,linkcolor=blue,urlcolor=blue]{hyperref}
\usepackage[usenames,dvipsnames]{xcolor}
\usepackage{gensymb}
\usepackage{float}
\usepackage[T1]{fontenc}
\usepackage[latin9]{inputenc}

\begin{document}

\title{\textcolor{MidnightBlue}{
Quantum anomalous Hall phase in  synthetic bilayers via twistless twistronics }}
\author{Tymoteusz Salamon}
\affiliation{ICFO - Institut de Ciencies Fotoniques, The Barcelona Institute of Science and Technology, Av. Carl Friedrich Gauss 3, 08860 Castelldefels (Barcelona), Spain}
\author{Ravindra W. Chhajlany}
\affiliation{Faculty of Physics, Adam Mickiewicz University, 61614 Poznan, Poland}
\author{Alexandre Dauphin}
\affiliation{ICFO - Institut de Ciencies Fotoniques, The Barcelona Institute of Science and Technology, Av. Carl Friedrich Gauss 3, 08860 Castelldefels (Barcelona), Spain}
\author{Maciej Lewenstein}
\affiliation{ICFO - Institut de Ciencies Fotoniques, The Barcelona Institute of Science and Technology, Av. Carl Friedrich Gauss 3, 08860 Castelldefels (Barcelona), Spain}
\affiliation{ICREA, Pg. Lluis Companys 23, Barcelona, Spain}
\author{Debraj Rakshit}
\affiliation{ICFO - Institut de Ciencies Fotoniques, The Barcelona Institute of Science and Technology, Av. Carl Friedrich Gauss 3, 08860 Castelldefels (Barcelona), Spain}
\affiliation{Max-Planck-Institut f{\"u}r Quantenoptik, D-85748 Garching, Germany}

\date{\today}
\begin{abstract}

We recently proposed quantum simulators of "twistronic-like" physics based on ultracold atoms and  synthetic dimensions [Phys. Rev. Lett. 125, 030504 (2020)]. Conceptually, the scheme is based on the idea that a physical monolayer optical lattice of desired geometry is upgraded to a synthetic bilayer system by identifying the internal states of the trapped atoms with synthetic spatial dimensions. The couplings between the internal states, i.e. between sites on the two layers, can be exquisitely controlled by laser induced Raman transitions. By spatially modulating the interlayer coupling, Moir{\'e}-like patterns can be directly imprinted on the lattice without the need of a physical twist of the layers. This scheme leads practically to a uniform pattern across the lattice with the added advantage of widely tunable interlayer coupling strengths. The latter feature facilitates the engineering of  flat bands  at larger "magic" angles, or more directly, for smaller unit cells than in conventional twisted materials. 
In this paper we extend these ideas and demonstrate that our system exhibits topological band structures under appropriate conditions. 
To achieve non-trivial band topology we consider imanaginary next-to-nearest neighbor tunnelings that drive the system into a quantum anomalous Hall phase. 
In particular, we focus on three groups of bands, whose  their Chern numbers triplet can be associated  to a trivial insulator (0,0,0),  a standard non-trivial (-1,0,1) and a non-standard non-trivial (-1,1,0). 
 We identify regimes of parameters where these three situations occur. 
We show the presence of an anomalous Hall phase and  the appearance of topological edge states. Our works open the path for experiments on topological effects in twistronics without a twist.
\end{abstract}

\maketitle
\textcolor{MidnightBlue}{\section{Introduction}}

Twistronics (from twist and electronics) is a term commonly used nowadays to describe  the physics resulting from the twist between layers of two-dimensional materials. This terminology was introduced in Ref.~\cite{Carr2017}, which conducted theoretical studies on how a twist between the layers can change electonic properties of bilayer graphene. But the history of this new area of research goes back to Ref.~\cite{Castro-Neto2007}, whose authors suggested that twisted bilayer graphene could provide a new material with unprecedented properties. Flat bands at the magic angle were discovered in 2011~\cite{Barticevic2010}, whereas Bistritzer and MacDonald showed that for a twisted material with a ``magic angle'' the free electron properties radically change~\cite{Allan2011}. More recently, two seminal experimental papers~\cite{Cao18-1, Cao18-2} demonstrated that such twisted bilayer graphene at the magic angle can host both strongly insulating Mott states and superconductivity. These impressive results triggered an avalanche of experimental and theoretical works~\cite{Yankowitz19,Lu19,Volovik18,Yuan18,Koshino,Ochi,Zou18,Peltonen18,Padhi18,Sboychakov19,Guinea18,Xu18,Wu18,Isobe18,You18,Lian19,Lin19,Song19} (See Ref.~\cite{Balents2020} for a recent review article). Many of these recent activities discuss topological insulators in magic-angle twisted bilayer graphene and the possibility of creating and controlling topological bands in these systems~\cite{Park19,Song19,Ma19}.

\begin{figure}[t!]
\centering
\includegraphics[clip,width=1\columnwidth]{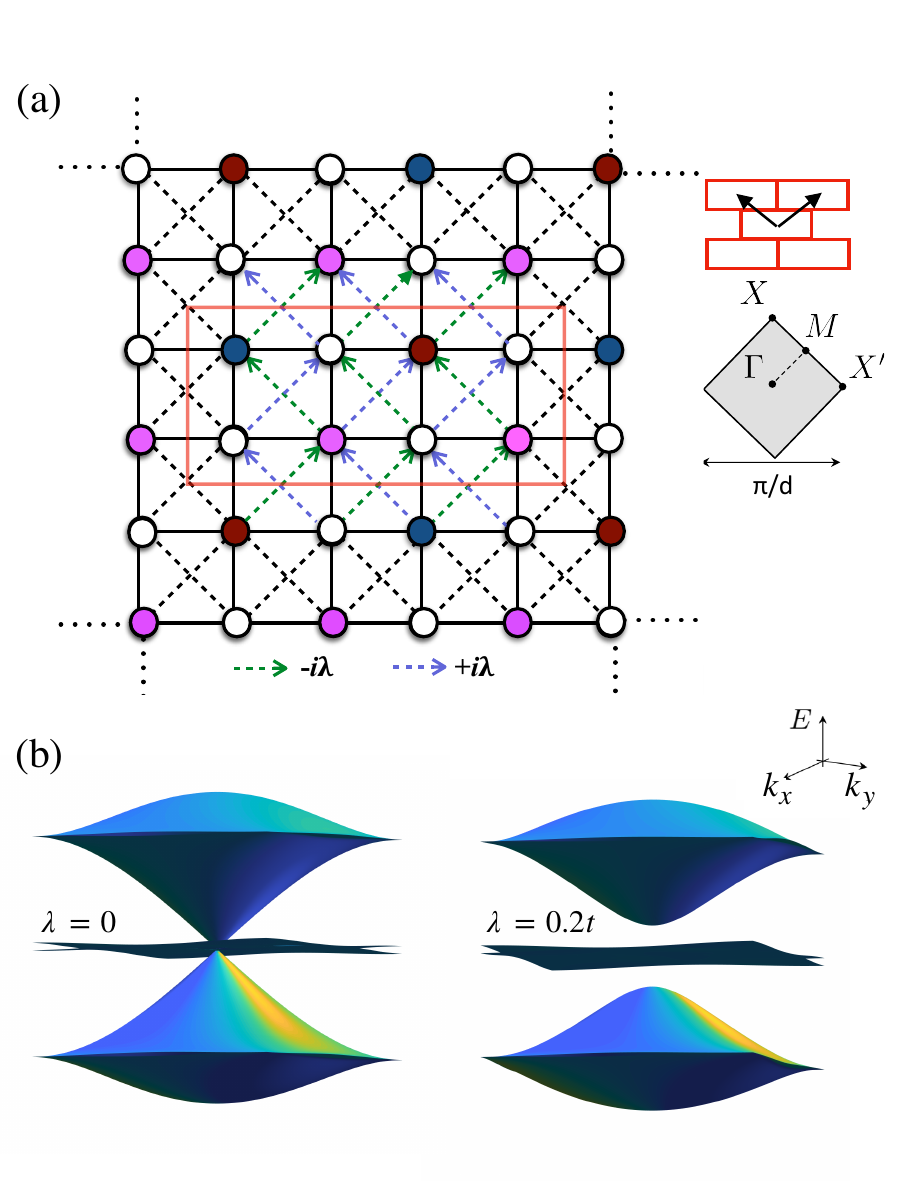}
\caption{{The supercell structure and effects of complex hopping.} (a) The left panel illustrates a sketch of a single plane of the bilayer structure corresponding to one of the spin states $m=\pm 1/2$. We consider square lattice potential. The maroon and dark-blue sites are Raman coupled with the second plane under the chosen spatial modulation of the synthetic tunneling. The pink sites (in both the layers) experience a chemical potential, $\mu$. The real-space nearest-neighbor tunneling $t$ is shown by the solid lines. The next-nearest-neighbor complex tunneling depending on directions of hopping and the position of the lattice sites is shown by staggered dotted lines. The red line shows a top view of the elementary unit cell of the system containing $2 \times 8 $ sites for Raman periodicities $l_x=l_y=4$, where the factor of $2$ accounts for the two layers. 
The right top panel depicts the arrangement of elementary unit cells and the two translation vectors as arrows. This leads to the first Brillouin zone shown in the right bottom panel, indicating also the position of the high-symmetry points. (b) The three dimensional view of the six band manifold in the vicinity of $E=-\Omega_0(1-\alpha)$ for $\Omega_0 \alpha/h=40 t$ with $\alpha=0.2$, $\gamma=\pi/2$ and $\mu=0$. The left panel shows the case, when next-nearest-neighbor complex hopping is absent, i.e., $\lambda=0$.
It has two quasi-flat bands at the Dirac points of two dispersive bands in form of Dirac cones. The right panel shows the same set of bands, but in presence of the complex tunneling, $\lambda=0.2t$. This causes opening of a hard gap between the quasi-flat bands and the nearby dispersive bands.}
\label{Fig1}
\end{figure}

Topological order has now become a central research topic in physics, exemplified by the 2016 Nobel Prize for D. J. Thouless, F.D.M. Haldane, and J.M. Kosterlitz~\cite{Kosterlitz_2017,Haldane_2017}. The intimate relation between topology and condensed matter goes back to the discovery of the Integer Quantum Hall Effect (IQHE)~\cite{von_klitzing_86}: a 2D electron gas at low temperature and under a strong magnetic field presents a quantized transverse conductivity very robust against local perturbations. It was soon realized~\cite{thouless_82} that this robustness was coming from a new paradigm: a global topological order which cannot be described by the usual Ginzburg Landau theory of phase transitions. In the particular case of the IQHE, the presence of a strong magnetic field results in the appearance of flat bands (Landau levels), each of them being characterized by a distinct topological invariant, called Chern number, and the transverse conductivity is equal to the sum of the Chern numbers of the occupied Landau levels. Soon after, F.D.M. Haldane proposed the quantum anomalous Hall effect, which presents a quantized transverse conductivity but no Landau levels~\cite{haldane_88}. Such a toy model turned out to be the crucial ingredient for the original proposal of topological insulators in graphene~\cite{Kane05,Kane05_2} and stimulated very rapid progress of the area of topological/Chern insulators, topological superconductors, topological flat bands, and even systems with higher-order topology~\cite{Hasan10,Xi11,Benalcazar17,Suzuki05,Sato17,Yao12}. One of the most challenging and still persisting questions are related to the role of interactions, in particular strong interaction and correlations~\cite{Rachel_2018}. Interestingly, the interactions do not always destroy the topological phases. Strong interactions in flat band topological materials can lead to the fractional quantum Hall effect~\cite{Laughlin_99,Tsui_99,Stormer_99} or to fractional Chern insulators~\cite{Regnault_11,Yao_13,Liu_13,Grushin_15}. Furthermore, strong interactions can induce topology through a spontaneous symmetry breaking mechanism as it is the case in the celebrated topological Mott insulator~\cite{Raghu_2008,Sun_08,Dauphin_12,grushin2011,grushin2013,Zeng_18,Sur_18,Julia_20}.

Novel insights into the physics of topological order can be provided by quantum simulators. These highly flexible experimental systems are used to mimic complex quantum systems in a clean and controllable environment. Quantum simulators constitute one of the four major pillars of contemporary quantum technologies~\cite{AcinRoadmap}, and can be realized with various platforms such as ultracold atoms, trapped ions, superconducting qubits, circuit QED,  Nitrogen vacancies in diamond, or nanostructure in condensed matter (for a review see~\cite{Bloch08,specialissueNaturePhys,specialissueNaturePhys1,specialissueNaturePhys2,specialissueNaturePhys3,specialissueNaturePhys4}). In this work, we focus on the simulation of twisted graphene with ultracold atoms in optical lattices ~\cite{Lewenstein12}. Such platforms allow one to simulate diverse geometries leading to, among others, graphene-like physics in synthetic hexagonal and brickwall lattices~\cite{Soltan-Panahi11, Tarruell12, Jo12} or flat band physics in the Lieb lattice~\cite{Taie15,Ozawa17}.  There are indeed other platform for realizing hexagonal lattices~\cite{Polini}, but atoms combine additional advantages,  as explained herafter.  Cold atoms provide a unique playground for synthetic gauge fields~\cite{JakschZoller, Goldman_2014,reviewsongaugefields}, combined with time dependent lattice modulations/Floquet engineering~\cite{Eckardt,reviewonfloquet}. In particular, such techniques led to the experimental realization of the Hofstadter model~\cite{Monika1,Monika2}, the Haldane model~\cite{Jotzu,Hamburg,Asteria_2018}. Furthermore, artificial gauge fields can also be engineered with the help of synthetic dimensions which allows to engineer topological insulators on ladders with both non trivial Chern numbers and topological edge states~\cite{Boada12,Celi13,Stuhl_2015,Mancini1510,Mugel2017,Genkina_2019,Ozawa19}. Finally, optical lattices offer possibilities to study bilayer systems and proximity effects~\cite{Grass16}. 
\\
The rapid development of twistronics in condensed matter physics of 2D material stimulated extensive quest for quantum simulators of twistronics with ultracold atoms~\cite{Tudela19},  and of Moir\'e patterns in photonic systems~\cite{lluis2019}. We combined these two worlds in a recent work, quantum simulators and synthetic dimensions, and proposed twistronics without a twist~\cite{Salamon19}. In this paper we considered a single 2D optical lattice with a desired geometry (honeycomb, brick, or $\Pi$-flux square), and created multilayer sytems employing internal state of the atoms inserted in the lattice. These could be in the simplest case fermions with spin 1/2 or 3/2. The moir\'e patterns were generated by spatial modulations of the Raman  transitions coupling the internal states. Since the strength of the Raman coupling can be efficiently controlled in this system, the appearance of flat bands is expected to occur for much larger ``magic'' angles, or better to say for elementary cells of  much more modest sizes, like unit cell consisting of only $2\times 8$ lattice sites. We analyzed the properties of the band structure in such systems,  and showed that  these expectations were indeed correct.
In the present paper we develop further the idea of Ref.~\cite{Salamon19} and demonstrate that such system is very flexible and can be tweaked to exhibit topological band structure in various situations. In particular, to achieve non-trivial topology we engineer artificial complex next-to-nearest neighbor tunneling analogous to the ones appearing in the Haldane model~\cite{haldane_88}. Typically, the energy bands of our interest in this system form three groups and the Chern number changes from trivial (0,0,0), to a topological phase with a trivial flat band (-1,0,1) and a topological phase with a non-tivial flat band (-1,1,0). We identify the regimes of parameters where these three situations occur, and study properties of the system with periodic and open boundary conditions.
The paper is organized as follows. In Section~\ref{sec:system} we present the details of the model~\cite{Salamon19}, together with modifications required for achievement of  topological bands. In Section~\ref{sec:magic_angle} we discuss magic configurations and quasi-flat bands. Section~\ref{sec:SCH} is devoted to the investigations of the effects of onset of staggered hoping and  appearance of topological insulators. Similar effects and topological properties are discussed for dimerized lattices in Section~\ref{sec:DH}. Section~\ref{sec:exp} contains a short discussion of  feasibility of experimental realization of the discussed physical effects. The conclusions and outlook are presented in Section~\ref{sec:conlusions}.

\textcolor{MidnightBlue}{\section{The system}\label{sec:system}}
We consider a system of synthetic spinfull fermions in a bilayer material. The fermions are subjected to a synthetic magnetic field which leads to a  supercell structure. We propose the following scheme to realize such Hamiltonian in a quantum simulator. We consider ultracold fermionic atoms with four internal states, ${m,\sigma}=\pm1/2$, in a two-dimensional spin-independent optical square lattice with lattice spacing $d$. The four internal states are chosen such that two spin flavours of electrons are described by two pairs of the internal states, denoted by $\sigma$. The pair of spin states corresponding to the same $\sigma$ are subjected to the Raman coupling. The index $m$, which  distinguishes the Raman coupled pairs, is the labelling of the synthetic dimension. The two possible values of $m$ effectively realize a synthetic bilayer structure. A moir\'e-like supercell structure can be obtained by modulating the Raman coupling strength, $\Omega(x,y)$, in space. The complete Hamiltonian reads

\begin{equation}
\label{eq:Ht}
H = H_t+H_\lambda+H_\Omega+H_\mu, 
\end{equation}
where 
\begin{equation}
\begin{split}
H_t =& -\sum_{{\bf r},m,\sigma} t({\bf r}) \left[a_{m,\sigma}^{\dagger}({\bf r}+\mathbf{1}_x) + a_{m,\sigma}^{\dagger}({\bf r}+\mathbf{1}_y) \right]a_{m,\sigma}({\bf r})\\
&+\mathrm{h.c.}
\end{split}
\end{equation}
is the nearest neighbor hopping Hamiltonian with a real and space dependent tunneling amplitude $t(\mathbf{r})$,
\begin{equation}
\begin{split}
H_\lambda &= \sum_{{\bf r},m,\sigma} \lambda \left[\exp(i \phi_R(\vec{r})) a_{m,\sigma}^{\dagger}({\bf r}+\mathbf{1}_x+\mathbf{1}_y) + \right. \\
& \left. \exp\left(i \phi_L(\vec{r}) \right)  a_{m,\sigma}^{\dagger}({\bf r}-\mathbf{1}_x+\mathbf{1}_y)\right] a_{m,\sigma}({\bf r})+\mathrm{h.c.}
\end{split}
\end{equation}
is the the next-to-nearest hopping Hamiltonian with a complex tunneling amplitude $\lambda$ and a staggered phase $\Phi$,
\begin{equation}
H_\Omega = \sum_{{\bf r},m,\sigma} \Omega({\bf r}) \exp(-i { \bm \gamma}\cdot {\bf r})~a_{m+1,\sigma}^{\dagger}({\bf r})a_{m,\sigma}({\bf r}) +\mathrm{h.c.}
\end{equation}
denotes the synthetic hopping Hamiltonian with a space dependent Raman coupling $\Omega$ and a magnetic phase $\bm{\gamma}=\gamma(\mathbf{1}_x+\mathbf{1}_y)$, and 
\begin{equation}
H_\mu = \sum_{{\bf r},m,\sigma} \mu({\bf r})~a_{m,\sigma}^{\dagger}({\bf r}) a_{m,\sigma}({\bf r}),
\end{equation}
is the onsite chemical potential Hamiltonian with an amplitude $\mu(\mathbf{r})$. Figure~\ref{Fig1} provides a schematic depiction of the system under study.

The spatial modulation of the Raman coupling is chosen to be $ \Omega(x,y) = \Omega_0 \left[1 -\alpha(1+\cos{(2 \pi x/l_x)} \cos{(2 \pi y/l_y)}) \right]$, where $l_x$ ($l_y$) is its periodicity along the $x$ ($y$) axis.  In the following, we consider two distinct cases: \\
\textbf{(i) Staggered complex hopping (SCH).} we set $t({\bf r}) = t$ and fix the phases associated with the next-nearest neighbor complex tunneling amplitude by setting $\phi_L({\bf r})-\phi_R({\bf r})=\pi$, where  $\phi_R({\bf r} )=(2 \, {\bf r} .\mathbf{1}_y+1)\pi /2$ and $\phi_L({\bf r} )=(2 \, {\bf r} .\mathbf{1}_y+3)\pi /2$.\\
\textbf{(ii) Dimerized lattice (DL).} We consider dimerized real tunneling amplitude, such that $t({\bf r})$ takes the form of $t_1$ and $t_2$ in alternative sites in the $x$ and $y$ directions, along with the next-nearest neighbor complex hopping, where we set  $\phi_R({\bf r})=\phi_L({\bf r})=\pi/2$.

\textcolor{MidnightBlue}{\section{Magic configurations and quasi-flat bands}\label{sec:magic_angle}}

\begin{figure}[t!]
\centering
\includegraphics[clip,width=0.74\columnwidth]{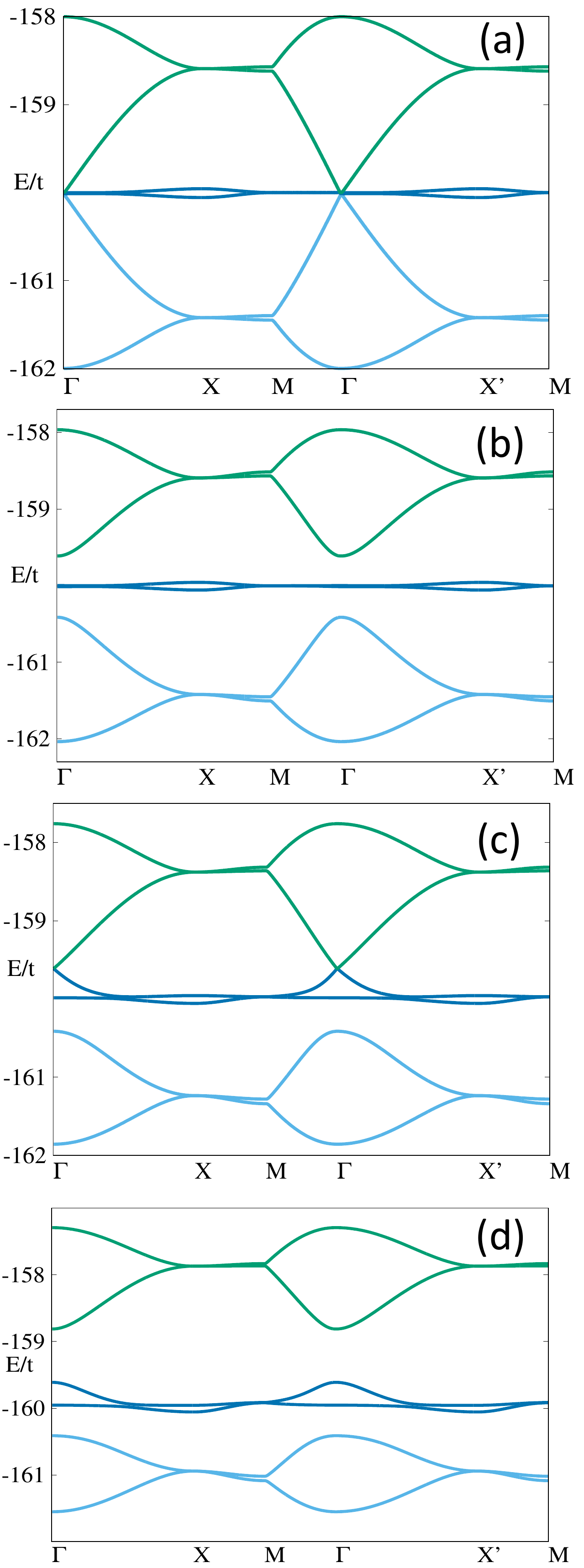}
\caption{{Magic configuration band structures in presence of staggered complex hopping.} Band structures corresponding to $\Theta(4,4)$ supercell along the paths passing through the high-symmetry points ${\bf \Gamma, X, M, \Gamma, X', M}$ for $\Omega_0 \alpha/h=40 t$, $\alpha=0.2$, and $\gamma=\pi/2$. Panel (a) shows the the set of six-bands around energy $-\Omega_0 (1-\alpha)$ with  $\lambda = 0.0$, while other bands, which are well separated (by atleast the energy of $\Omega_0 \alpha$) are not shown. There is no hard gap between the two middle quasi-flat bands and adjacent dispersive Dirac cones. Panels (b-c) reveal evolution of the spectrum with respect to the chemical potential, $\mu$. Panels (a), (b) and (c) corresponds to $\mu= 0$, $0.4t$ and $1.2t$, respectively. Finite $\lambda$ induces a hard gap between the quasi-flat bands and dispersive bands. The chemical potential leads the system through a non-trivial gapped-gapless transition, given the Fermi energy is adjusted accordingly.}
\label{Fig2}
\end{figure}

In the following, we study the Hamiltonian~\eqref{eq:Ht} under various boundary conditions: (i) periodic boundary conditions in both spatial ($x$ and $y$) direction, (ii) periodic boundary condition in one of the spatial direction (here along $x$), and (iii) open boundary conditions in all spatial directions. We first consider the case (i), for which the quasimomentum ${\bf k}=(k_x,k_y)$ is a good quantum number. In this case, we can apply the Bloch theorem, along with a gauge transformation, such that $a_{m,\sigma}({\bf r})=\sum_{\bf k} \exp\left( i({\bf k}\cdot{\bf r}+m{\bm \gamma}\cdot{\bf r})\right) a_{m,\sigma}({\bf k})$. The Hamiltonian can then be rewritten as $H=\sum_{{\bf k}} H_{{\bf k}}$ and can be diagonalized. The spatial periodicity of the synthetic tunneling fixes the supercell dimension, and hence the dimension of $H_{\bf k}$. The notation $\Theta(l_x,l_y)$ is introduced to represent corresponding supercell configuration.

The band structure corresponding to the case with $\lambda=0$ and $\mu=0$ has been studied in detail in the previous work. Remarkably, $\Theta(4 \nu,4 \nu)$ configurations, i.e., when $l_x=l_y=4\nu d$, with $\nu$ integer, were identified as magic configurations. $\Theta(4 \nu,4 \nu)$ configurations host quasi-flat bands surrounded by dispersive Dirac cone spectra with controllable Dirac velocities, and hence share certain characteristics associated with magic angle twisted bilayer graphene. In this work we focus on the the smallest possible configuration $\Theta(4,4)$ consisting of only 16 lattice sites. The corresponding Brillouin zone of the bilayer system showing the high-symmetry points is depicted in Fig.~\ref{Fig1}(a). In the following we briefly review few consequences for the case with $\lambda=0$ and $\mu=0$.

In the strong Raman coupling limit $(\Omega_0 \alpha)  \gg t$, isolated sets of narrow spin degenerate bands appears at the energies $\pm\Omega_0$, $\pm\Omega_0(1-\alpha)$ and $\pm\Omega_0 (1-2\alpha)$. We identify the energy spectrum of the system to be symmetric around zero energy, i.e. $E(\vec{q})=-E(\vec{q})$ and therefore we restrict our discussion to the negative energy bands. A non-isolated set of six bands around the energy $E/t=-\Omega_0(1-\alpha)$ are of particular interest in this work. Figure~1(b) shows the energy  sptectrum for a representative case with $\Omega_0 \alpha/h=40 t$, $\alpha=0.2$, $\lambda=0$, $\mu=0$ and the flux $\gamma=\pi/2$ (also see Fig.~\ref{Fig2}(a), which depicts the energy spectrum along the paths passing through the high-symmetry points). Here the parameters are chosen by considering experimentally accessible regime in practice. Within this six band manifold, the top and bottom three bands are symmetric around the energy $-\Omega_0 (1-2\alpha)$.  Noticeably,  two middle bands, which are quasi-flat, are formed closest to $-\Omega_0(1-\alpha)$ and they are sandwiched between dispersive bands in form of Dirac cones. These well isolated six band manifold with a total dispersion of $\Delta_6 = 4 \sqrt{2} t \cos(\gamma/2) + O(t^2/(\Omega_0 \alpha))$ are separated from remaining nearby bands by an energy gap of $~\Omega_0 \alpha$. The flatness of the quasi-flat bands (two middle bands) can be tuned precisely by adjusting the Raman coupling strength. Their approximate width can be derived within the second order perturbation theory as
\begin{eqnarray}
    \Delta_F = \frac{t^2 \cos ^2\left({\gamma}/{2}\right)}{\Omega_0\alpha} \left(\frac{24\alpha^3-88\alpha^2+106\alpha-32}{3\alpha^3-11\alpha^2+12\alpha-4}\right).
 \end{eqnarray}
The relative flatness of the bands is defined as $F=\Delta_F/\Delta_6$. As a result, the relatively flatter middle bands can be obtained by increasing $\Omega_0 \alpha$. The  parameters here are chosen from experimentally accessible regime. It is worth mentioning that the Dirac velocity is proportional to $\cos(\gamma/2)$, and hence the bandwidths of the dispersive Dirac bands can be controlled separately by tuning $\gamma$.

Within these six bands, the system has no gaps. The upper dispersive Dirac cone touches the middle quasi-flat bands at the high symmetry point $\Gamma$, i.e., $k_x=k_y=0$. A  non-zero flux, however, opens a tiny \emph{local} gap between the lower dispersive bands and the middle quasi-flat bands at $\Gamma$. Finite flux breaks rotational and time reversal symmetries, $c_4\tau$, of the system. In the following we discuss strategies for opening a gap around the Dirac band touching and the resulting topological phases of matter.

\textcolor{MidnightBlue}{\section{Staggered complex hopping} \label{sec:SCH}}

\textcolor{MidnightBlue}{\subsection{Bulk properties of the system}}

\begin{figure}[t!]
\centering
\includegraphics[clip,width=1\columnwidth]{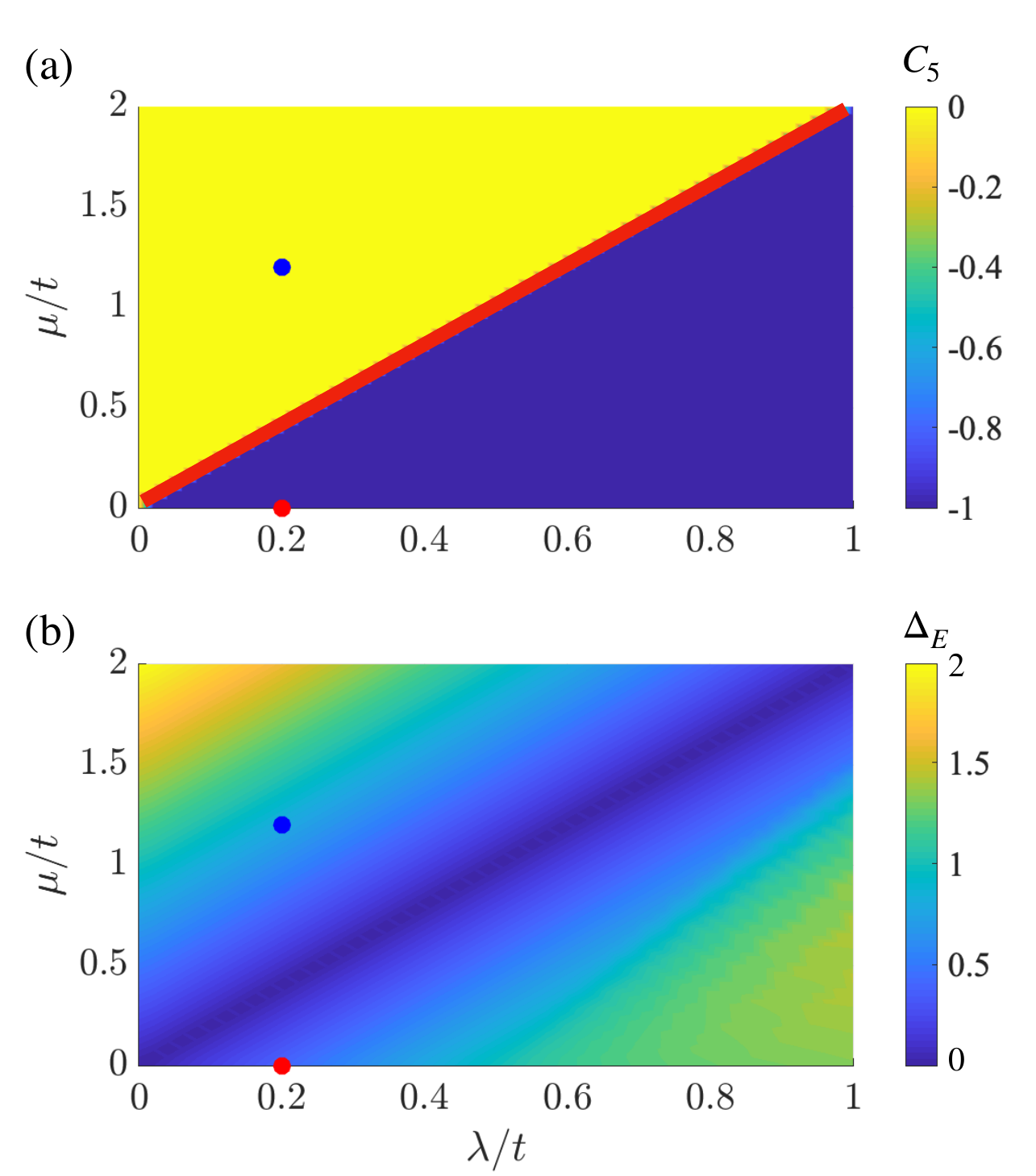}
\caption{
(a) Chern number of the five lowest bands as a function of system' parameters $\lambda /t$ and $\mu /t$. As seen on (b) the system is gapless for $\mu = 2\lambda$, which makes $C_5$ undefined in this region. The gapless region visible on (b) is marked by the red line on (a). Topological transition presented on the Fig.\ref{Fig2} can be seen by looking at the vertical cross-section at $\lambda/t=0.2$. Blue region of chern number -1 indicates the configurations at which system has standard non-trivial topology since $C_3=-1$ and the order is $(-1,0,1)$, while yellow area depicts parameters' values for which non-standard topological order $(-1,1,0)$ can be obtained. Red and blue points mark the values of the parameters for which edge states have been plotted on Fig.\ref{Fig4} (a) and (b),respectively.
(b) Second band gap (energy gap between the hybridized quasi-flat middle bands and the dispersive upper band) as a function of $\lambda/t$ and $\mu/t$. Vanishing energy gaps mark a topological phase transition in the system. Meaning of red and blue points remains the same as on the panel (a). }
\label{Fig2p1}
\end{figure}

We investigate the staggered complex hopping case [see Eq.~\eqref{eq:Ht} and Fig.~\ref{Fig1}(a)].  We first set $\mu=0$ and consider periodic boundary conditions in both $x$ and $y$ directions. A finite value of $\lambda$ breaks time-reversal symmetry. We focus on the same six-band subset of the energy spectrum. A non-zero value of $\lambda$ induces a mass term at the Dirac crossing. As a result, three isolated sets of bands are formed - each of them consists of two hybridized bands. The resulting band structure along the high-symmetry points is shown in Fig.~\ref{Fig2}(b) for $\Omega_0/h=200 t$, $\alpha=0.2$, $\gamma=\pi/2$, and $\lambda = 0.2t$. While the quasi-flat bands remain closest to the energy $\Omega_0(1-\alpha)$, the top (bottom) dispersive bands shifts upwards (downwards). The band gaps can be controlled by tuning the value of $\lambda$. As we will see, such hopping amplitude $\lambda$ drives the system into a quantum anomalous Hall phase~\cite{haldane_88}.

We further investigate the influence of a non-zero staggered chemical potential $\mu$ on the system. A Finite value of $\mu$ breaks the inversion symmetry of the system and the energy spectrum is no longer exactly symmetric around the zero energy. Nevertheless, the overall qualitative features of the negative and positive energy bands remain closely similar for a value of $\mu$ smaller than the central bandgap. Moreover, this staggered chemical potential results in a significant asymmetry between the two top and two bottom bands within the six band manifold under study. As we discuss below, this staggering potential has a prominent impact on topological phases of the system. Figures~\ref{Fig2} (b-d) shows the evolution of the band structure for increasing values of $\mu$: the band gap between the top dispersive bands and the middle quasi-flat bands shrinks, and the gap closes at $\mu_c\simeq2\lambda$. For $\mu>\mu_c$, the gap then reopens. Figure~\ref{Fig2p1} shows the energy gap between the set of quasi-flat bands and the upper set of bands. We also observe a gap closing and reopening of the lower gap for increasing staggering potential with opposite sign. The latter is reminiscent of the interplay between the staggering potential and the imaginary next-to-nearest neighbor hopping in the Lieb lattice~\cite{weeks_2010}. In fact, the system undergoes a topological phase transition. We characterize the topology of the system with the help of the Chern number $\mathcal{C}$, a topological invariant for the class of Chern insulators. The Chern number of the $n$-th band is defined as
\begin{equation}
\begin{split}
    \mathcal{C}_n &= i\int_{BZ} \mathit{F}^{xy}_{n} ({\bf k})dS \\
    &= i\sum^{N_xN_y}_{l=1}\int_{P_l}\mathit{F}^{xy}_{n,l} ({\bf k})dS
\end{split}
\end{equation}
the integral of the Berry curvature $\mathit{F}^{xy}_{n,l}({\bf k})=\nabla_{{\bf k}} \times \mathit{A}^n({\bf k})$ written in terms of the Berry connection $\mathit{A}^n({\bf k})=\bra{u^n({\bf k})}\partial_{{\bf k}}\ket{u^n({\bf k})}$. In the second equality, we have rewritten this integral  as a sum of integrals over the plaquettes of the discretization grid of the Brillouin Zone.  For a such a discretized grid, the Berry curvature can be be computed numerically with the help of the FHS algorithm~\cite{Suzuki05}
\begin{equation}
  \mathcal{C}_n \approx \frac{1}{2\pi i}\sum_{P_l}\Big{(}\braket{u^n_k}{u^n_m}\braket{u^n_m}{u^n_o}\braket{u^n_o}{u^n_p}\braket{u^n_p}{u^n_k}\Big{)},
\end{equation}
where $P_l$ denotes a plaquette in the Brillouin Zone with four vertices $(k,m,o,p)$ labeled in the anti-clockwise order with k being top left vertex and $\ket{u^n_k}$ is a Bloch function corresponding to the n-th eigenvalue at point $k$. The summation is performed over all plaquettes in the Brillouin Zone. We emphasize that for degenerate bands one has to use the algorithm proposed in Ref.~\cite{Hatsugai_1993} to compute the non-abelian Berry curvature. In this manuscript, we use the total Chern number 

\begin{equation}
C_i = \sum_{j=1}^i \mathcal{C}_j
\end{equation}

defined as the sum of the Chern numbers of the $i$ first occupied bands.

As can be seen in Fig.~\ref{Fig2p1}(a), for $\mu < \mu_{c}$, The hybridized middle set of bands are topologically trivial with zero Chern number, while the bottom (top) set of dispersive band is topologically nontrivial with $C=$-1(1).  The gap closing leads to a topological phase transition with a transfer of Chern number from the upper set of bands to the middle set of bands~\cite{Belissard_95}, and for $\mu>\mu_c$, the middle set of bands becomes non trivial with $C=1$ and the upper set of bands becomes trivial with  $C=0$.  The bottom dispersive bands remain non-trivial with $C=-1$. In the following subsections, we discuss in detail the topological edge states appearing in a system with boundaries.

\textcolor{MidnightBlue}{\subsection{Cylindrical geometry and edge states}}

\begin{figure}[t!]
\centering
\includegraphics[clip,width=1\columnwidth]{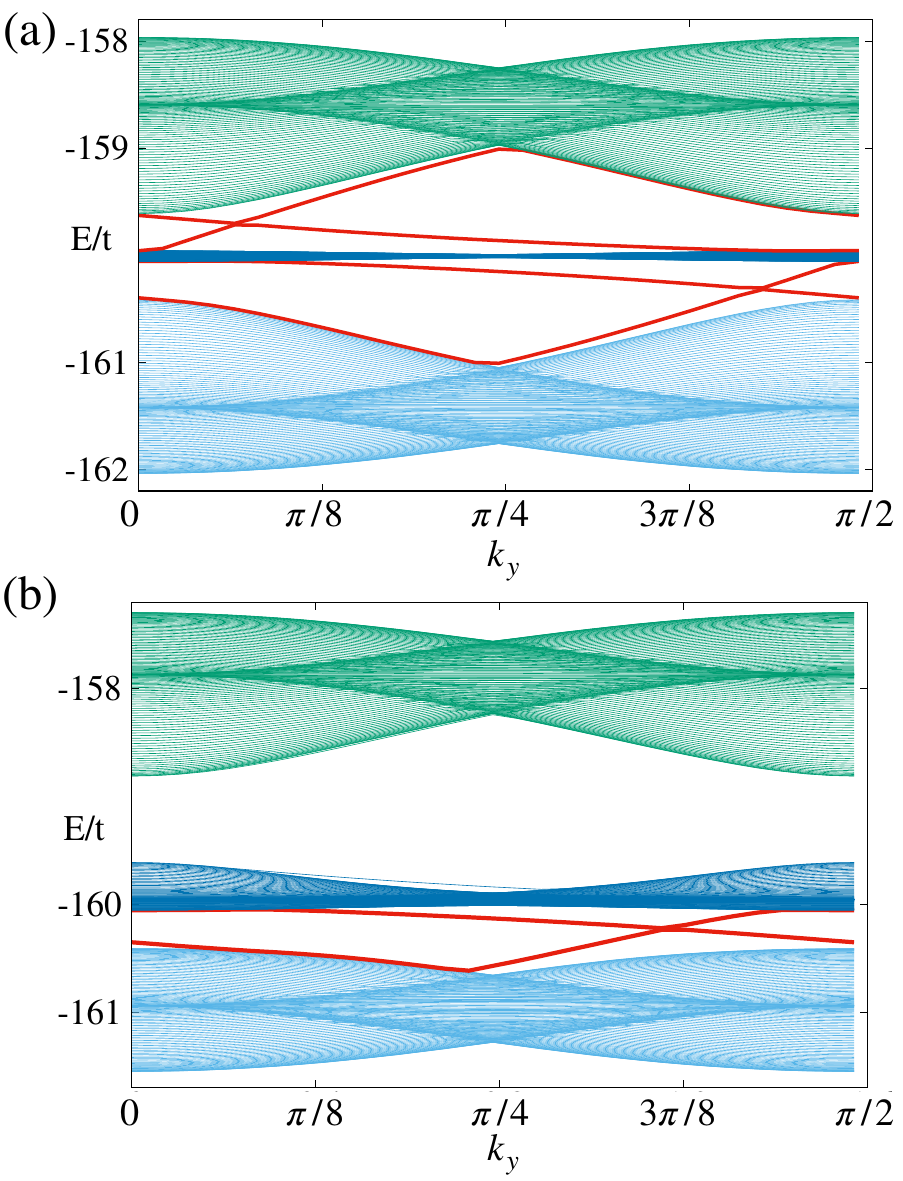}
\caption{{Edge and bulk dispersions for cylindrical geometry.} Energy spectrum is shown to demonstrate two distinct cases with (a) $\mu=0$ and (b) $\mu=1.2$ for the parameters $\Omega_0 \alpha/h=40 t$, $\alpha=0.2$, $\gamma=\pi/2$, $\lambda = 0.2t$. The mid-gap states due to open boundaries along the $x$-axis are shown by red lines. These mid-gap states are edge states connecting the energetically separated bulk bands.}
\label{Fig3}
\end{figure}

In order to study the topological edge states of the system,  we consider a cylindrical or strip geometry.  Calculations are conveniently performed using an enlarged unit cell consisting of plaquettes of size $4\times 4$ site per layer. More specifically, we consider a bilayer strip of finite length along the $x$ direction and  infinite length in the $y$ direction through periodic boundary conditions. As a result $k_y$ remains a good quantum number. 
The enlarged unit cell is repeated $n_x$ times in the $x$ direction. 

\begin{figure}[t!]
\centering
\includegraphics[clip,width=0.85\columnwidth]{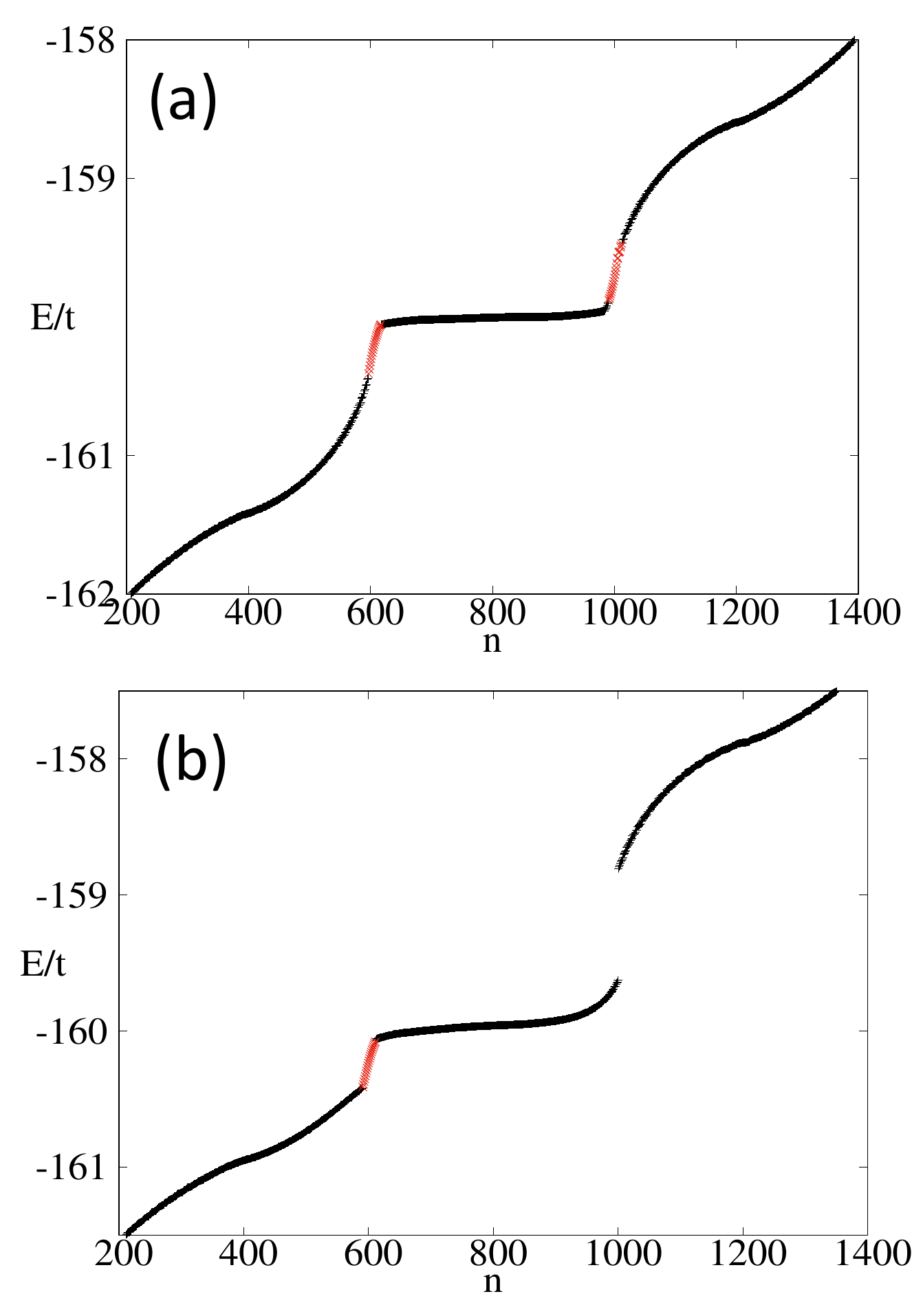}
\caption{Energy spectrum of the synthetic bilayer square lattice with $2 \times 40 \times 40$ sites and open boundary condition for (a) $\mu=0$ and (b) $\mu=1.2t$ for the parameters $\Omega_0=200$, $\alpha=0.2$, $\gamma=\pi/2$ and $\lambda=0.2t$. The bulk states are shown by black plus symbols. The new mid-gap states appearing due to open boundary condition is shown by red crosses. These are the edge states. For (a)$\mu=0$, the edge states appear in both the energy gaps.  However, for $\mu>\mu_{c}$ (shown here (b)$\mu=1.2t$, the edge states appear only within the first energy gap. }
\label{Fig4}
\end{figure}
\begin{figure}[t!]
\centering
\includegraphics[clip,width=0.82\columnwidth]{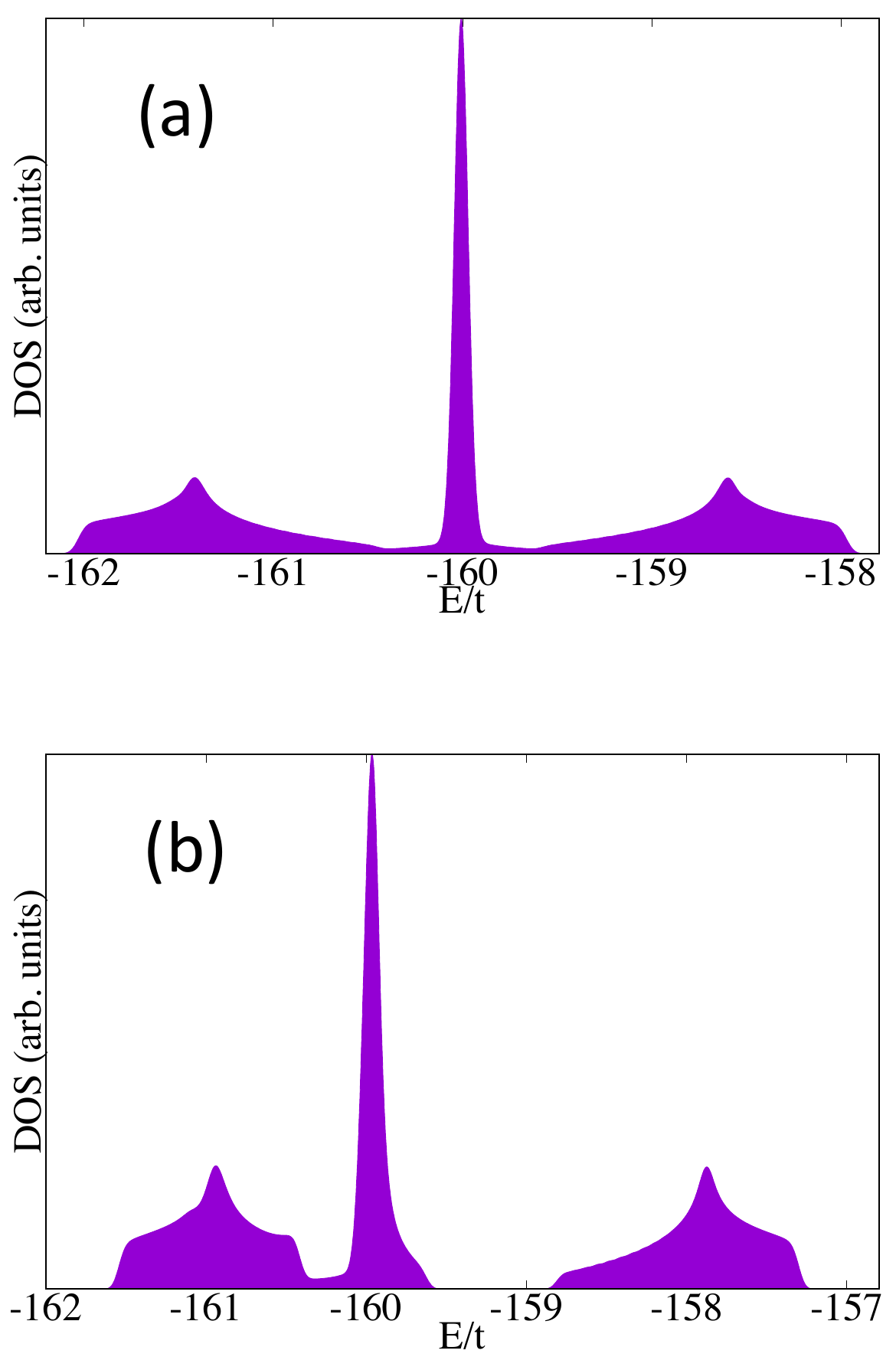}
\caption{DOS (in arbitrary units) of the synthetic bilayer square lattice with $2 \times 40 \times 40$ sites and open boundary condition for (a) $\mu=0$ and (b) $\mu=1.2t$ for the parameters $\Omega_0=200$, $\alpha=0.2$, $\gamma=\pi/2$ and $\lambda=0.2t$.}
\label{Fig5p2}
\end{figure}

In the following, we fix the set of parameters as in the previous section, i.e. $\Omega_0/h=200 t$, $\alpha=0.2$, $\gamma=\pi/2$, and $\lambda = 0.2t$,  and we concentrate on the same energy window close to the energy $-\Omega_0(1-\alpha)$. Figure~\ref{Fig3} shows the energy spectrum as a function of the quasi momentum $k_y$ for $n_x=30$. 
Figure~\ref{Fig3}(a) depicts the energy spectrum for $\mu=0$ [red dot in Fig.~\ref{Fig2p1}(b)]: The system presents three energy bands that were already present in the previous section, which we call bulk energy bands. Interestingly, the bulk energy gaps host topological edge states in accordance to the celebrated bulk-edge correspondence~\cite{Hatsugai_1993}. These topological edge modes are responsible for edge currents and their number at each edge is equal to the sum of Chern numbers of the occupied bands.  Hence, setting the Fermi energy $E_F$ at one of the two energy gaps, one finds one edge state at each edge, which is, as expected. Figure~\ref{Fig3} (b) shows the energy spectrum for $\mu=1.2t>\mu_c$ [blue dot in Fig.~\ref{Fig2p1}(b)]. In this case the first (bottom) energy gap supports topological edge states, while the second energy gap does not. This is again consistent with the fact that for this case the bottom band has $\mathcal{C}=-1$, and hence there is one edge state at each edge in the first gap, while the sum of Chern numbers of bottom and middle bands is zero. Hence depending on the choice of $E_F$, we can have a Chern insulator insulator or a trivial insulator. In order to achieve a closer understanding of the edge states and the bulk-boundary correspondence, we further analyze a finite square lattice in the following discussion.

\textcolor{MidnightBlue}{\subsection{Finite square lattice and edge states}}

\begin{figure}[t!]
\centering
\includegraphics[clip,width=1\columnwidth]{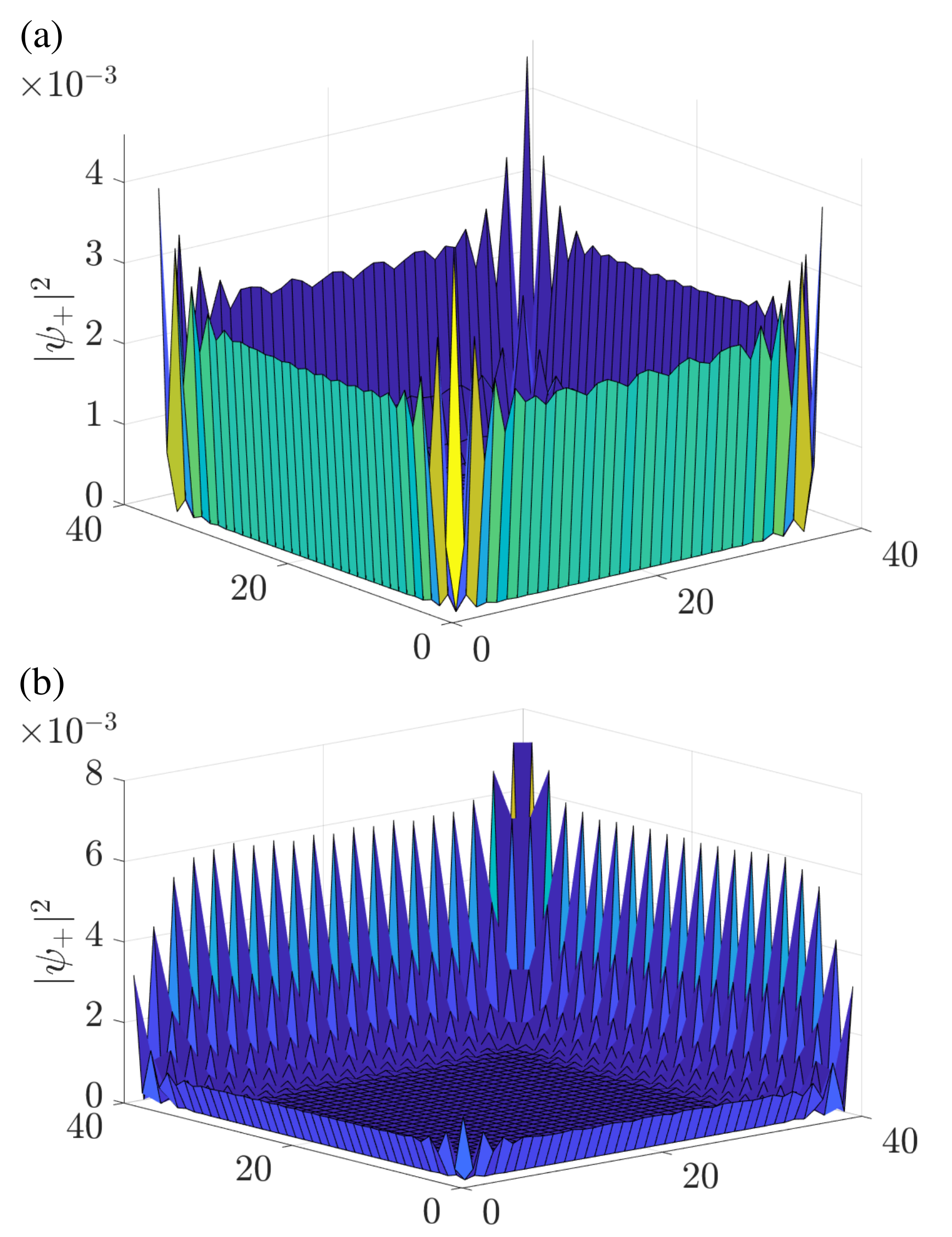}
\caption{Probability density plots for an edge mode corresponding to energy $E\approx-160.28 t$. (a) The wave function is fully localized on the edges of the lattice, which in this case has a length of 39 sites. (b) The wave function is localized on two edges of the lattice which contain raman coupled sites. In this case has a length of 40 sites, commensurate with the periodicity of the unit cell.  }
\label{Fig5}
\end{figure}

We finally consider a bilayer finite square lattice with $2 \times L \times L$ sites.  We do not impose periodic boundary condition, i.e., both the layers are open in both $x$ and $y$ directions. Unlike previous cases, the system can not be associated with a good quantum number due to the absence of any periodicity. We solve the Hamiltonian in Eq.~(\ref{Fig1}) by diagonalizing the matrix with $2L^2 \times 2L^2$ entries. We again focus on the energy bands close to $-\Omega_0(1-\alpha)$ (see Figs.~\ref{Fig2} and \ref{Fig3}). Figure~\ref{Fig4} shows the sorted eigenvalues of the Hamiltonian for a system of length $L=40$ and for the same set of parameters of the previous subsections. We again show two different topological phases for $\mu=0$ [Fig.~\ref{Fig4}(a)] and $\mu=1.2t$ [Fig.~\ref{Fig4}(b)].

The appearance of the new states due to the absence of periodicity are shown by red crosses. These states are detached from the bulk states and clearly are a manifestation of the open boundaries of the layers. These are edge states and, as we demonstrate below, live on the boundaries of the layers. For $\mu<\mu_{c}$, the edge states appear in both energy gaps. This is consistent with the discussions for the cylindrical geometry and the computation of the bulk Chern numbers. This is illustrated in Fig.~\ref{Fig4}(a) as an exemplary case with $\mu=0$. For $\mu>\mu_{c}$, as expected, the first energy gap hosts new states, while the second energy gap does not. Figure~\ref{Fig4}(b) demonstrates this via an example with $\mu=1.2t$. The corresponding density of states (DOS) are shown in Fig.~\ref{Fig5p2}.  For $\mu<\mu_{c}$, two regions with low density of states appear adjacent to $-\Omega_0(1-\alpha)$ under open open boundary condition. These correspond to the energies of the mid-gap states in the energy gaps of the bulk spectrum. For $\mu>\mu_{c}$,  one of these two regions has vanishing DOS due to absence of the mid-gap states between the middle band and the upper dispersive band.

In order to characterize the edge states in real space, we define probability density of the edge state corresponding to $m=\pm 1$ , $p_{\pm}({\bf r}) = \abs{\braket{\psi^E_{\pm}({\bf r})}{\psi^E_{\pm}({\bf r})}}^2$, where $\psi^E_{\pm}({\bf r})$ denotes the eigenvector whose energy is the closest to $E$ the and $m=\pm 1$ denotes the projection on one of the synthetic dimensions. Interestingly, the edge states have a different real space profile depending on the commensurability of the number of sites with the supercell: lattices with the length commensurate with the periodicity of the unit cell have two edges with Raman coupled sites and two edges that do not contain such sites [see Fig.~\ref{Fig1}(a)], while lattices with the length $\mod(\frac{L}{\Theta})=\pm 1$ have all 4 edges of the same kind. Therefore to maintain the symmetry of the edges, where we expect the wave function to localize, system depicted on Fig.~\ref{Fig5}(a) was decreased by one site in each direction to $L_{x,y}$=$39$ resulting in symmetric probability distribution over all 4 edges of the lattice. On the other hand, for the lattice lengths being a integer multiple of unit cell's periodicity the edge states are also localized on the borders of the lattice, but the probability density is not equally distributed, favouring two edges, which are not Raman coupled. Such behaviour can be observed on Fig.~\ref{Fig5}(b), which shows the spatial density distribution, $p({\bf r})$, of a typical edge states in the topologically insulating phases of the system with $L$=$40$. We have verified that $p_{-}({\bf r})$ exhibits similar features. The chosen edge state is a mid-gap state in the first band gap for $\mu=0$. The spatial distribution has an asymmetric nature, which comes from the finite size synthetic bilayer  geometry governed by the interlayer coupling pattern. The edge states are more localized at two adjacent edges of lattice corners, which host alternative sites with Raman coupled internal states, and are rather weakly localized in rest of the boundaries, where internal states in any of the sites are not subjected to such Raman coupling by construction. \\

\textcolor{MidnightBlue}{\section{Dimerized lattice}\label{sec:DH}}

We now focus on the DL case, which is based on the alternating NN tunneling both in x and y direction and a complex NNN hopping. 

.In particular we consider both dimerized NNN hopping, as in the SCH case as well as unstaggered NNN tunneling with $\phi_R = \phi_L$ that provides a zero net flux per palquette. First, we analyze the possibility of using dimerization as a substitute for space dependent chemical potential for obtaining the non-trivial gap between dispersive and quasi-flat sets of bands. Second, we simplify the NNN hopping leaving the dimerization untouched, since realizing constant diagonal hopping experimentally is less complicated.

The effect of the lattice dimerization has been primarily studied in Ref.~\cite{Brown540}. The asymmetry of the hopping leads to the shift of the bands in the energy spectrum together with energy gap openings. It allows for the isolation of the set of quasi-flat bands, which originally are not separated by the global gap from the rest of the spectrum (see Fig.~\ref{Fig8}(a)). However, this gap opening is trivial and leads to the triplet of Chern numbers $(0,0,0)$. 

\begin{figure}[t!]
\centering
\includegraphics[clip,width=1\columnwidth]{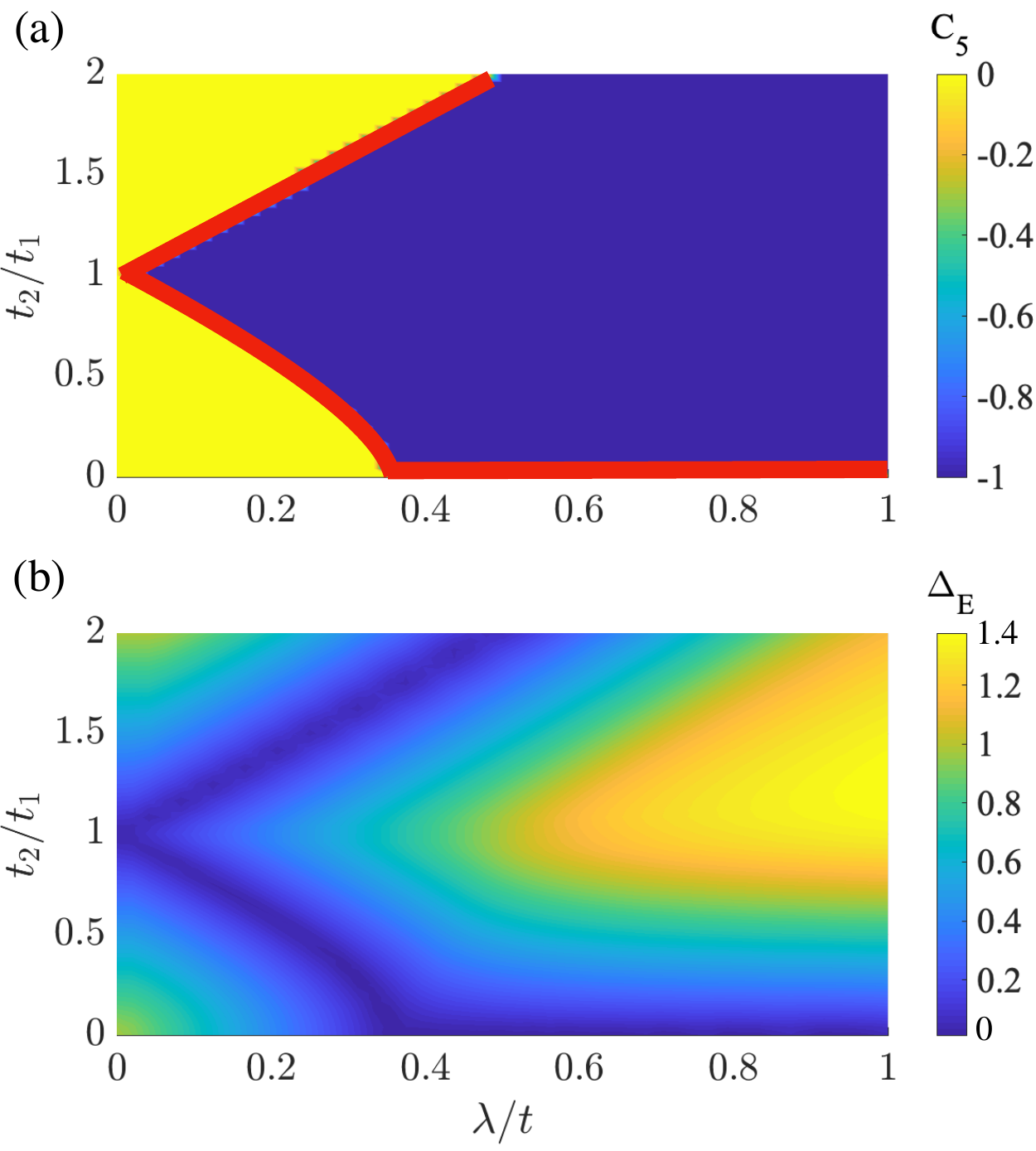}
\caption{Chern number of the lower dispersive set of bands (Chern number of the set of quasi-flat bands is always 0). In the absence of the staggered NNN hopping the topology is trivial, regardless the strength of dimerization( this case is depicted by the first column of the plot). Increasing the NNN hopping causes the change of the topological order into the standard nontrivial one after reaching gapless phase, which is visible on (b) and marked by the red line on (a).  The order of the signs of the Chern numbers depends on the dimerization sign. Moreover for $t_2/t_1=0$ discussed set of bands consists of 3 separate subsets of bands with trivial topology( this case is depicted by the lowest row of the plot). Dispersion of each of these subsets depends on the value of $\lambda$ and for $\lambda=0$ all 3 subsets are almost perfectly flat.}
\label{Fig7}
\end{figure}

\begin{figure}[t!]
\centering
\includegraphics[clip,width=1\columnwidth]{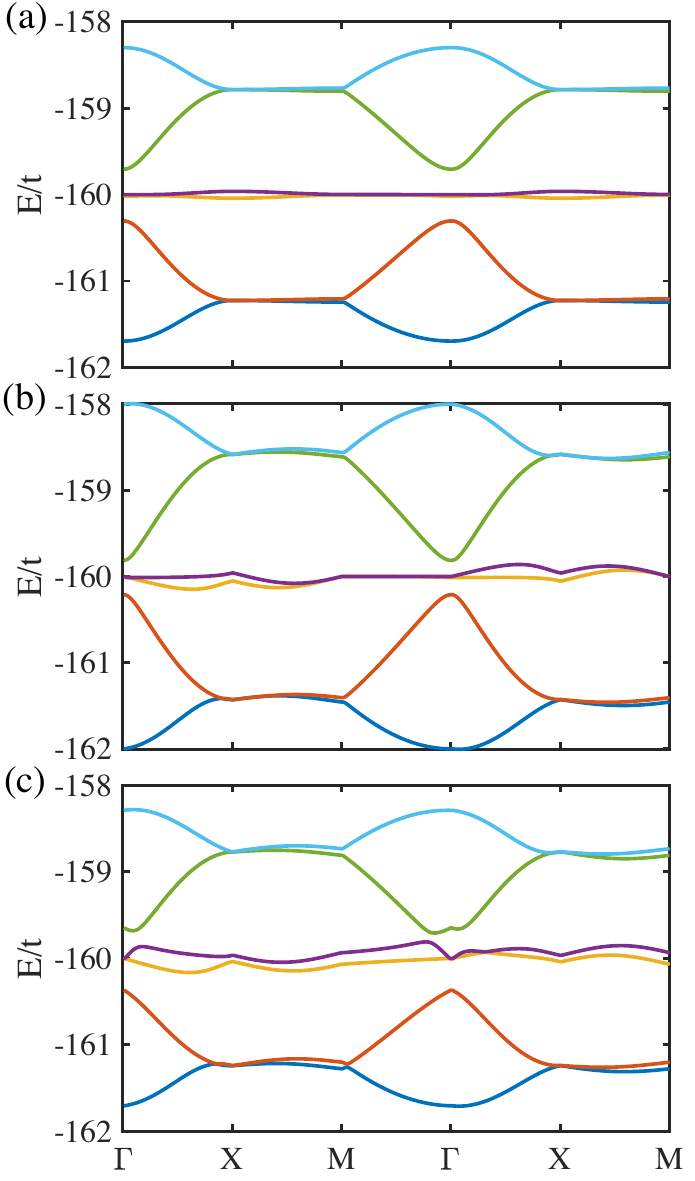}
\caption{Magic configuration band structure in presence of uniform complex NNN hopping.
(a) $t_2/t_1=0.7$ and $\lambda=0$ shows the effect of pure dimerization of the lattice in the absence of NNN hopping resulting in vanishing Dirac cones.  In comparison with Fig.\ref{Fig2}(b), which represents staggered NNN hopping without chemical potential one can see that dispersion of the quasi-flat bands is much bigger.
(b) $t_2/t_1=1$ and $\lambda=0.2$ represents the spectrum of the system with uniform NNN complex hopping generating the  the increased dispersion of the quasiflat bands and the lack of the global gap and.
(c) In this case $\lambda = 0.2$ $t_2/t_1=0.7$. Dispersion of the quasiflat bands is further increased but lattice dimerization provides the global gap. However these two effect compete and the system can be gapless for differernt values of $\lambda$ (see Fig~\ref{Fig9}(b)). }
\label{Fig8}
\end{figure}

\paragraph{Staggered NNN hopping} The interplay of the NN dimerization and the staggered NNN hopping results in a topological phase diagram depicted Fig.~\ref{Fig7}. Such order was also obtained in the SCH case in absence of chemical potential. Hence, one can conclude that dimerization does not affect order-changing processes but allows one to observe the edge states of the well separated bands.

\begin{figure}[t!]
\centering
\includegraphics[clip,width=1\columnwidth]{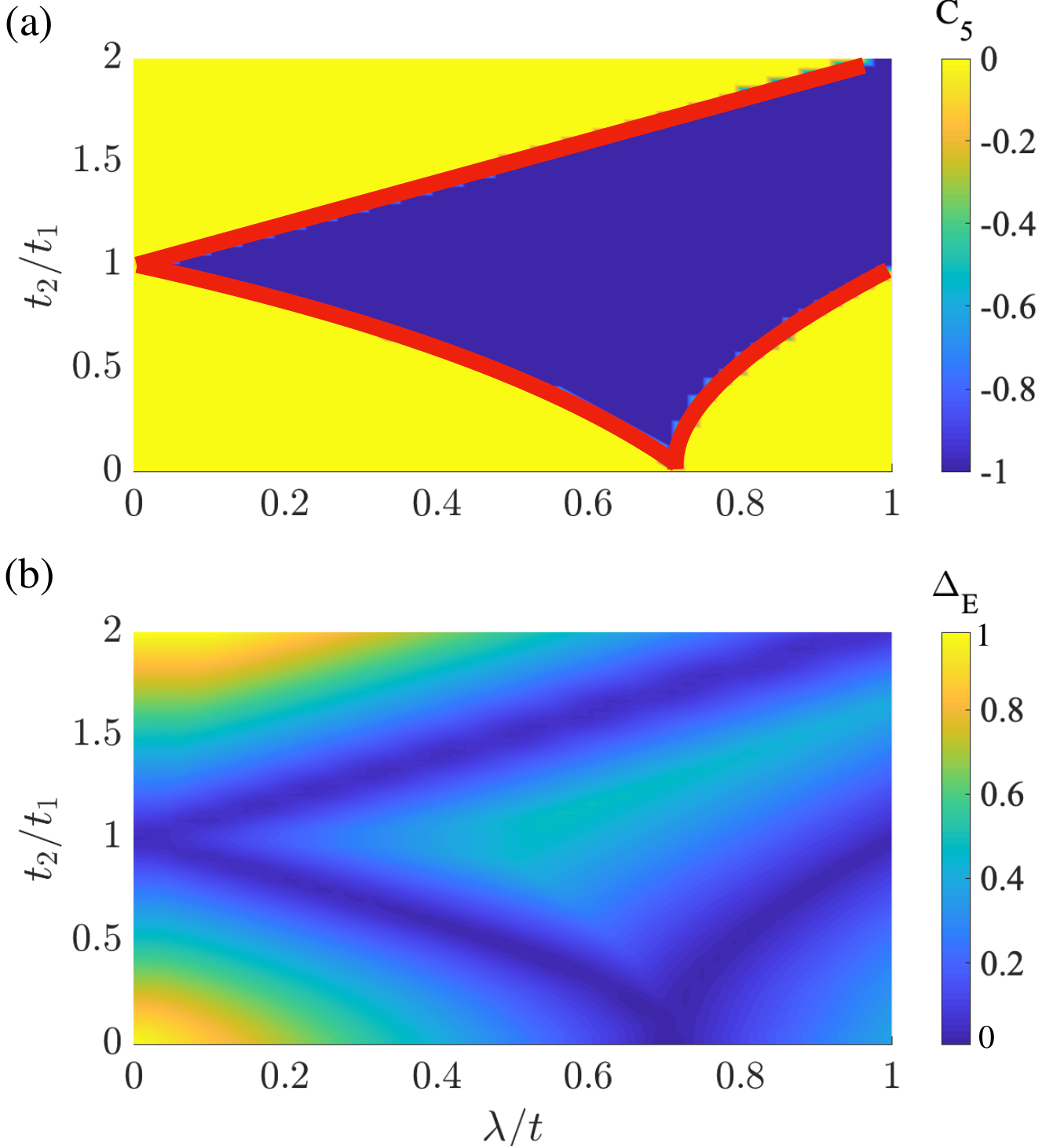}
\caption{Chern number of the lower dispersive set of bands (Chern number of the set of quasi-flat bands is always 0). In the absence of the uniform NNN hopping the topology is trivial, regardless the strength of dimerization( this case is depicted by the first column of the plot). Increasing the NNN hopping causes the change of the topological order into the standard nontrivial one. However the order of the signs of the chern numbers depends on the dimerization sign. Moreover for $t_2/t_1=0$ discussed set of bands consists of 3 separate subsets of bands with trivial topology( this case is depicted by the middle row of the plot). Dispersion of each of these subsets depends on the value of $\lambda$ and for $\lambda=0$ all 3 subsets are almost perfectly flat.}
\label{Fig9}
\end{figure}

\paragraph{Uniform hopping} We now consider dimerized lattice with uniform NNN hopping within the single layer. As discussed in the previous paragraph, in this approach the opening of a global gap is guaranteed by the dimerization of NN hopping. The role of complex NNN tunneling is more complicated to explain, since unlike staggered NNN it cannot be associated with Hall phase nor opens a global gap between quasi-flat and dispersive bands, as shown on Fig.\ref{Fig8}(b). On the other hand, similarly to Ref.~\cite{Kane05} in such case the net flux per plaquette is 0. All above suggests that applying diagonal (NNN) hopping of the complex value does not imply topological non-triviality. Indeed uniform NNN hopping does not open the global gap under any parameters' configuration. On the other hand, similarly to the effects caused by dimerization in the absence of NNN hopping, previously quasi-flat bands become dispersive with increasing $\lambda$ together with the shift in xy-plane. All effects of the uniform complex NNN tunneling can be seen on the Fig.~\ref{Fig8}(c). The effect of dimerized NN hopping is similar to band separation observed in staggered NNN hopping except, of course, the topological order of the system with uniform NNN tunneling varies between the trivial and standard non-trivial with no possibility of reaching non-standard topology, as presented in Fig.~\ref{Fig9}.

\textcolor{MidnightBlue}{\section{Experimental scheme}\label{sec:exp}}

We here discuss a  quantum simulation scheme of the system. As we will see below, all elements of the proposed scheme have been sucessfully implemented in state-of-the-art experiments. The challenge consists in combining all the necessary ingredients. We proceed in this section in steps, with the message directed mostly to the experimentalists.  First, we review the basic scheme for twistronics without a twist, proposed already in Ref.~\cite{Salamon19}. Then, we discuss the necessary additional ingredients of both proposed schemes:  I) Staggered complex hopping, and II) dimerized real tunnelling. We discuss possible methods that can be used to realize our schemes: i) Laser induced tunneling; ii) Floquet engineering; and iii) Super-lattice and holographic potential imprinting methods. Finally, we discuss possible detection schemes of the topological properties of the model. \\

\subsection{Basic experimental scheme} 

We proposed in Ref.~\cite{Salamon19} to use a subset of four states out of the large nuclear spin manifold $I=9/2$ of $^{87}$Sr, or $^{173}$Yb ($I=5/2$). The $SU(N)$ invariance inhibits collisional redistribution of the atoms among the different states. We select two of them to be $\sigma=\,\,\uparrow$, and the other two to be  $\sigma=\,\,\downarrow$. All are subjected to a two-dimensional spin-independent optical lattice potential, created by two counter-propagating lattice beams. We choose the laser wavelength $\lambda_L=813$ nm (corresponding to the magic wavelength of the clock transition $^1S_0\rightarrow\,^3P_0$). We set a lattice depth to about 8 recoil energies, $8\,E_L$, which yields tunneling of order of 100 Hz. Lattice constant as usual is $d=\lambda_L/2$.

To create the synthetic layer tunneling, we exploit two-photon Raman transitions between spins $m=\pm 1/2$. A pair of Raman beams with $\lambda_R=689$ nm near-resonant to the intercombination transition  $^{1}S_0\rightarrow\, ^{3}P_1$, would produce a coupling of amplitude $\Omega_0=\Omega_1\Omega_2/\Delta_0$. Here $\Omega_1$ and $\Omega_2$ are the individual coupling amplitudes of the Raman lasers and $\Delta_0$ the single-photon detuning. The Raman beams propagate in a plane perpendicular to the lattice potential, are aligned along its diagonal, and form an angle $\theta$ with the lattice plane (see Fig. 1 in ref. \cite{Salamon19}). This yields an in-plane momentum transfer per beam $k_R=\pm2\pi \cos{\theta}/\lambda_R$, with projections $k_R/\sqrt{2}$ along the lattice axes. Therefore, the phase of the synthetic tunneling is $\bm{\gamma}\cdot{\bf r}=\gamma (x \hat{x}+ y \hat{y})$, with $\gamma=\pm2\pi \cos{\theta} \lambda_L/(\sqrt{2}\lambda_R)$. The sign is determined by the relative detuning of the Raman lasers. Experimentally, the simplest choice is to use counter-propagating Raman beams ($\theta=0\degree$), which yields $\gamma=0.8$ (mod $2\pi$). However, other magnetic fluxes can be easily realized by adjusting the value of $\theta$.

To implement a periodic modulation of the Raman coupling amplitude on the scale of several lattice sites, we propose to exploit a periodic potential created by a laser close-detuned from the excited state to excited state transition $^{3}P_1\rightarrow\,^{3}S_1$ (corresponding to $688$ nm \cite{Stellmer13}). This results in a large light shift of the $^{3}P_1$ excited state of amplitude $\delta$, leading to a detuning of the Raman beams $\Delta (x,y)=\Delta_0+\delta (1+\cos{(2\pi x/l_x)} \cos{(2\pi y/l_y)})$. Its effect is to modulate the Raman coupling amplitude $\Omega(x,y)\simeq \Omega_0[(1-\alpha) - \alpha \cos{(2\pi x/l_x)} \cos{(2\pi y/l_y)}]$, with $\alpha=\delta/\Delta_0\sim 0.2$ for realistic experimental parameters \cite{Stellmer13, Chen19}. We therefore named it ``modulation laser''.

\subsection{Extensions of the basic scheme}

\textbf{SCH case.} Here the NN tunnelling is standard and constant, $t({\bf r}) = t$.  The staggered chemical potential is given by $\vec{\mu}=\mu (\hat{x}+\hat{y})$, $\hat{n}=2d (\hat{x}+\hat{y})$, and can be relatively easily realized using super-lattice or holographic potential imprinting methods. The challenging part here is relate to the phases associated with the next-nearest neighbor complex tunnelings, set to  $\phi_L(\vec{r})-\phi_R(\mathbf{r})=\pi$, where  $\phi_R(\vec{r})=(2\mathbf{r}.\mathbf{1}_y+1)\pi/2$. We suggest to use here laser induced tunneling or lattice shaking. 
For laser induced tunneling one possibility would be to to employ the clock transition from $^1S_0 \to ^3P_0$, using appropriate
polarization of the assisting laser to couple to different excited states  (for instance, coupling $+3/2 \to + 1/2$ via $\sigma_-$ polarized light, coupling $+3/2 \to + 3/2$ via $\pi$-polarized light, and $+3/2 \to + 5/2$ via $\sigma_+$ polarized light). The main problem is that the clock transitions will cross-talk immensely with the light-shifting scheme of the $^3P_0$ state that we proposed to use to get the moir\'e pattern. In fact,  we should expect that the staggered complex hoppings will not only be staggered (if we design and realize the staggering well), but they will be spatially modulated as well. The period of the latter modulation should follow the period of our ``moir\'e'' pattern i.e.
\begin{equation}
    \lambda(\mathbf{r}) = \lambda_0+\Delta \lambda \cos{(2 \pi x/l_x)} \cos{(2 \pi y/l_y)}.
\end{equation}
After a careful study one observes however, that as long as $\Delta \lambda\sim t$, the effects of the spatial modulation of the NNN complex hopping remain marginal. In fact they are limited to negligible bandwidth corrections, which do not affect topological order nor open/close new gaps in the system.

\textbf{DL case.} In Eq.~\eqref{eq:Ht}, we considered dimerized real tunneling, such that $t({\bf r})$ takes the form of $t_1$ and $t_2$ in alternative sites in the $x$ and the $y$  directions, along with the next-nearest neighbor complex hopping. Here the situation seems to be easier from the experimental point of view. The alternating tunnelling can be achieved using the super-lattice techniques (dimerization). The next-nearest neighbor complex hopping with the homogenous phase set to $\phi_R(\mathbf{r})=\phi_L(\mathbf{r})=\pi/2$ should be accessible via lattice shaking and Floquet engineering techniques.   \\

\subsection{State-of-the-art experimental techniques}

\textbf{Laser assisted tunneling} The idea of employing laser assisted tunneling for generation of synthetic gauge fields goes back to the seminal paper of Jaksch and Zoller \cite{JakschZoller}.  It was generalized to non-Abelian fields in  Ref.~\cite{Osterloh2005}. These ideas all seemed very ``baroque'' at that time, but finally were realized in experiments with amazing effort, but equally amazing results~\cite{Monikasthesis,reviewsongaugefields,Monika1,Miyake,Monika2,Miyake}.

\textbf{Floquet engineering} In the context of cold atoms this technique goes back to the pioneering theory works of A. Eckardt and M. Holthaus~\cite{EWH2005}, followed by experiments of E. Arimondo and O. Morsch~\cite{Zenesini2009}. In condensed matter 
the works concerned  creation of topological phases in graphene~\cite{Oka2009,McIver2020}). The possibility of generating artificial gauge field was first discussed in Ref.~\cite{Eckardt} and realized in experiments of Hamburg group~\cite{Struck2011}. This culminated with the experimental realization of arbitrary complex phases~\cite{Struck2012}, and theoretical proposals combining shaking and on-site excitations~\cite{Hauke2012}. In the recent years many fascinating results were obtain using shaking (cf. \cite{Parker, Clark2018}, for a review see \cite{Eckardt2017}). In a sense, from the perspective of the present paper a culmination of these effort consisted in realization of the Haldane model with next nearest neighbor complex tunnelings in a brick lattice~\cite{Tarruell12,Jotzu}. Recently, the Hamburg group combined the studies of the Haldane model with the use of machine learning methods~\cite{Weitenberg2016,Tarnowski2017,Rem2019}.

\textbf{Super-lattice and holographic potential imprinting methods} These are nowadays standard methods, developed already many years ago and described in textbooks such as~\cite{Lewenstein12}. They have a plethora applications ranging from designing traps of special shape, through creation of dimerized  lattices, to imprinting random potentials. All of these methods ban be useful for our purposes, for instance for designing a  lattice with dimerized tunneling, etc. .  

\subsection{Detection of the topological order}
 
In cold-atom quantum simulators, the standard transport experiment techniques used to characterize the transverse conductivity in 2D materials are feasible but very demanding~\cite{Brantut_2012,Krinner_2014} and there is therefore the need for other detection schemes to characterize the topology of the system. In the last decade, many detection schemes have developed for quantum simulators~\cite{Goldman_2014,ozawa_19} and we briefly review here the a non-exhaustive list of techniques that could be applied to the synthetic twisted bilayer material. The total Chern number and the Berry curvature could be measured through the anomalous velocity of the center of mass of the atomic cloud~\cite{Price_2012,dauphin2013}. This technique, already applied in recent experiments~\cite{Aidelsburger_2014,Wintersperger_2020}, would require an additional optical gradient. Alternatively, the Chern number could measured through the depletion rate of the bands in the presence of heating~\cite{Tran2017}. This effect, called quantized circular dichroism, has been implemented in state-of-the-art experiment~\cite{Asteria_2018} and would require an additional shaking of the lattice. Finally, the topology could be characterized through the observation of the chiral edge states. The latter could be done by a suitable quench protocol~\cite{Goldman_2013b,Tran_2015}.

\textcolor{MidnightBlue}{\section{Conclusions - Outlook}\label{sec:conlusions}}

In the present paper we developed further the idea of ``twistronics without a twist'' and demonstrated that it can be used to engineer interesting topological band structures under various conditions.  Focussing on  a square lattice system with synthetic dimensions, we showed the appearance of an anomalous Hall phase in presence of  artificial  complex  next-to-nearest  neighbor  interlayer tunneling. Moreover, we discussed the emergence of topological bands via another mechanism --  lattice dimerization. In general,  the  bands  of interest can be categorized into three groups - trivial, and two categories  of topologically non-trivial  bands differing by their Chern number combinations: standard nontrivial, and non-standard not-trivial.

Possible directions of this research line in the near future concern  the incorporation of  interaction effects.   On the technical side,  in the first stage, incorporation of the interaction effects can be carried out via a mean field theory at the Hartree-Fock as well as ``slave boson/fractionalization'' level.  Moreover, as our scheme provides the possibility of observing physics similar to magic angle twisted bilayer graphene with an \emph{effectively} large rotation angle, implying a much smaller supercell, performing ab-initio calculation  could be possible via advanced tensor network algorithms. The pressing questions in these realm are: (i) the origins of strongly correlated in strongly correlated phenomena in twisted materials, such as unconventional superconductivity phenomena, (ii) the coexistence of superconducting and correlated insulating states in magic-angle twisted bilayer graphene, and their relationship, (iii)
the role of topology in the interacting systems, which can be probed by altering the quasiflatband topology.

\textcolor{MidnightBlue}{\section{acknowledgements}}

We thank Alessio Celi, Christoph Weitenberg, Klaus Sengstock and Leticia Tarruell for entlighting discussions. M.L. group acknowledges funding from European Union (ERC AdG NOQIA-833801), the Spanish Ministry MINECO and State Research Agency AEI (FIDEUA PID2019-106901GB-I00/10.13039 / 501100011033, SEVERO OCHOA No. SEV-2015-0522, FPI), European Social Fund, Fundaci\'o Cellex, Fundaci\'o Mir-Puig, Generalitat de Catalunya (AGAUR Grant No. 2017 SGR 1341, CERCA program, QuantumCAT$\_$U16-011424 , co-funded by ERDF Operational Program of Catalonia 2014-2020), MINECO-EU QUANTERA MAQS (funded by The State Research Agency (AEI) PCI2019-111828-2 / 10.13039/501100011033), and the National Science Centre, Poland-Symfonia Grant No. 2016/20/W/ST4/00314. 
T.S acknowledges additional support from the Secretaria d'Universitats i Recerca de la Generalitat de Catalunya and the European Social Fund, R.W.C.  from the Polish National Science Centre (NCN) under  Maestro Grant No. DEC-2019/34/A/ST2/00081, A.D. from the Juan de la Cierva program (IJCI-2017-33180) and the financial support from a fellowship granted by la Caixa Foundation (fellowship code LCF/BQ/PR20/11770012), and D.R. from the Fundaci\'o Cellex through a Cellex-ICFO-MPQ postdoctoral fellowship.

\bibliographystyle{apsrev4-1}


\end{document}